# Reverse Detection of Short-Term Earthquake Precursors
V. Keilis-Borok[1,2], P. Shebalin[1], A. Gabrielov[3], and D. Turcotte[4]

[1] *International Institute for Earthquake Prediction Theory and Mathematical Geophysics, Russian Ac. Sci., Warshavskoe sh., 79, korp. 2, Moscow, 113556, Russia*
[2] *Institute of Geophysics and Planetary Physics and Department of Earth and Space Sciences, UCLA, Los Angeles, CA, 90095-1567, USA*
[3] *Departments of Mathematics and Earth and Atmospheric Sciences, Purdue University, West Lafayette, IN 47907-1395, USA*
[4] *Department of Geology, University of California, Davis, Davis, CA 95616, USA*

Contacts: vkb@ess.ucla.edu, shebalin@mitp.ru

**Abstract**

1. Short-term earthquake prediction is dubbed the "Holy Grail of earthquake science." Here, we introduce a new approach to this problem, based on the concept of selforganization of seismically active fault network. That approach is named "*Reverse Detection of Precursors*" (*RDP*), since it considers precursors in reverse order of their appearance. First, we detect the "candidates" for the short-term precursors; in our case these are newly introduced chains of earthquakes reflecting the rise of an earthquake correlation range. Then we consider each chain, one by one, checking whether it was preceded by an intermediate-term precursor in its vicinity. If *yes*, we regard this chain as a precursor; in prediction it would start a short-term alarm. Such analysis has an essential advantage: the chain indicates the narrow area of possibly complex shape, where an intermediate-term precursor should be looked for. This makes possible to detect precursors undetectable by the direct analysis.

2. *RDP* can best be described on an example of its application; we describe retrospective prediction of two prominent Californian earthquakes - Landers, 1992, M = 7.6, and Hector Mine, 1999, M = 7.3, and suggest a hypothetical prediction algorithm. However, its validation is considered in subsequent studies, starting from (Shebalin et al, 2003a, b). The goal of this paper is to describe *RDP* methodology, since it has potentially important applications to many other data and to prediction of other critical phenomena besides earthquakes. In particular, it might vindicate some short-term precursors, previously rejected as giving many false alarms.

3. Validation of the algorithm per se requires its application in different regions with a substantial number of strong earthquakes. First (and positive) results are obtained for 21 more strong earthquakes in California (M ≥ 6.4), Japan (M ≥ 7.0) and the Eastern Mediterranean (M ≥ 6.5); these results are described elsewhere. The final validation requires, as always, prediction in advance for which this study sets up a base. We have the first case of a precursory chain reported in advance of a subsequent strong earthquake (Tokachi-oki, near Hokkaido island, Japan) Sept. 25, 2003, M = 8.1.

4. Possible mechanisms underlying *RDP* are outlined.

## 1. Introduction

Short-term earthquake prediction is dubbed the "Holy Grail of earthquake science" in the recent report of the US National Research Council (Jordan, 2003). It is indeed the most coveted and least accessible goal of earthquake prediction research. Here, we suggest a new approach to that problem, taking advantage of a fundamental feature of seismicity - its multi-scale selforganization.

**1.1. Generation of strong earthquakes – a non-localized process.** Seismicity is commonly recognized as a part of the geodynamics (Aki, 2003; Bird, 1998; Keilis-Borok, 1990; King et al, 2002; Press, 1965; Rundquist, 1999; Scholz, 1990); in seismically active areas the earthquakes accommodate a considerable fraction of tectonic development of the lithosphere. That development goes on in multiple time-, space-, and energy- scales and preparation of strong earthquakes is not an exception. Accordingly, while the target of earthquake prediction – a strong earthquake – is a localized event, the process of its generation is not localized. Strictly speaking, its time scales range from geological to seconds in time, and spatial scales - from global to microscopic (Turcotte, 1997, Keilis-Borok, 1990); however, in prediction research a truncated scaling is usually considered: from tens of years to days, and from hundreds of km to kilometer.

The multiplicity of scales is reflected in the general concept of the seismically active lithosphere as a hierarchical dissipative non-linear system, persistently self-organizing from time to time into the critical phenomena – the strong earthquakes (Blanter et al, 1997, Bowman et al, 1998; Gabrielov et al, 1994, 2000; Jaume et al, 1999; Keilis-Borok, 1990; Rundle et al, 2000; Sornette, 2000; Turcotte, 1997; Zaliapin et al, 2002a). Among manifestations of that selforganization are premonitory seismicity patterns – the spatio-temporal patterns of seismicity emerging as a strong earthquake approaches (Aki, 2003; Buffe et al, 1993; Caputo et al, 1983; Gabrielov et al, 1994; Jin et al, 2003; Keilis-Borok, 1990; 1996, 2000; Keilis-Borok et al, 1990a, b, 1964, 1999, 2002, Knopoff et al, 1996; Kossobokov et al, 1995, 2003; Ma et al, 1990; Mogi, 1985; Newman et al, 1995; Novikova et al, 2002; Press, 1965; Press et al, 1995; Romanowicz, 1993; Rotwain et al, 1999; Shebalin et al, 2000; Turcotte, 1997; Zaliapin et al, 2002a, b, 2003b; Zöller et al, 2001). A multitude of such patterns have been reported in rather different scales. Systematically tested are the intermediate-term patterns (with characteristic lead time of *years*). Here we suggest a method to detect the short-term patterns, which have the lead time of *months*.

**1.2. Reverse Detection of Precursors (*RDP*).** We consider the short-term patterns in conjunction with intermediate-term ones. This is done by *RDP* analysis, in which these patterns are detected in the reverse order of their appearance: short-term patterns are analyzed first, although they emerge later. Our findings can best be described on a specific example of data analysis.

**1.3. Region and data.** We describe detection of short-term patterns before two prominent Californian earthquakes - Landers, 1992, M = 7.6, and Hector Mine, 1999, M = 7.3. These are the largest Californian earthquakes since 1965 – the period, when the earthquake catalog is sufficiently complete for our analysis. Territory considered is shown in Fig. 1. The earthquake catalog is taken from (ANSS/CNSS and NEIC).

## 2. Chains

Our point of departure is provided by the short-term patterns *Roc* and *Accord* capturing a premonitory increase in earthquake correlation range. They were found first in models (Gabrielov et al, 2000) and then in observations (Keilis-Borok et al, 2002; Shebalin et al, 2000; Novikova et al, 2002). Other patterns capturing that phenomenon are suggested in (Zöller et al, 2001; Zaliapin et al, 2002b). Here, we introduce the pattern *Chain* which is a generalization of *Roc* and *Accord*. Qualitatively, a chain is a rapidly extended sequence of small earthquakes that follow each other closely in time and space.

### 2.1. Definitions.

*Earthquake catalog*. As in most premonitory patterns of that family (Keilis-Borok, 1996; Kossobokov et al, 2003) aftershocks are eliminated from the catalog; however, an integral measure of aftershocks sequence $b$ is retained for each remaining earthquake (main shocks and foreshocks). We use a common representation of the earthquake catalog $\{t_j, \varphi_j, \lambda_j, h_j, m_j, b_j\}$, j = 1, 2, … Here $t_j$ is the time of an earthquake, $t_j \geq t_{j-1}$; $\varphi_j$ and $\lambda_j$ – latitude and longitude of its epicenter; $h_j$ – focal depth; and $m_j$ – magnitude. We consider the earthquakes with magnitude $m \geq m_{\min}$. Focal depth is not used in this study.

*Chain*. Let us call two earthquakes "*neighbors*" if: (*i*) their epicenters lie within a distance $r$; and (*ii*) the time interval between them does not exceed a threshold $\tau_0$. A chain is the sequence of earthquakes connected by the following rule: *each earthquake has at least one neighbor in that sequence; and does not have neighbors outside the sequence*. The threshold $r$ is normalized by the average distance between the earthquakes with magnitudes $m \geq \underline{m}$, where $\underline{m}$ is the magnitude of the smaller earthquake in a pair considered. We use a coarse normalization $r = r_0 10^{c\underline{m}}$, $c$ being a numerical parameter.

Let $k$ be the number of earthquakes thus connected and $l$ - the maximal distance between their epicenters. We look for precursors only among the chains with $k \geq k_0$ and $l \geq l_0$. These thresholds ensure that our chains are exceptional phenomena.

*Chain's vicinity*. To compare location of a chain with locations of strong earthquakes we consider its *R*-vicinity formally defined as the union of circles of the radius $R$ centered at the epicenters of in the chains forming the chain. To smooth the borders of that area we add the dense sequence of circles along the lines connecting each epicenter in the chain with the two closest ones. The envelope of all the circles is the border of *R*-vicinity of the chain; it is similar to the "Wiener sausage", introduced by N. Wiener in the theory of probability.

**2.2. Data analysis**. We detected the chains defined above using numerical parameters listed in Table 1. Aftershocks have been identified by a coarse windowing, as described in (Keilis-Borok et al, 2002). The remaining catalog contains 3940 earthquakes. We have found among them nine chains, altogether containing 116 earthquakes: this shows that our chains are indeed exceptional phenomena. Maps of the chains are shown in Fig. 2; shaded areas are their vicinities, defined above. Vital characteristics of each chain are given in Table 2. Fig. 3 juxtaposes the chains and strong earthquakes on the time-distance plane; distance is counted along the dashed line shown in Figs. 1 and 2.

Table 1 Parameters for detecting the chains

| $m_{min}$ | $r_0$, km | $c$ | $\tau_0$, days | $k_0$ | $l_0$, km | $R$, km |
|---|---|---|---|---|---|---|
| 3.3 | 50 | 0.35 | 20 | 8 | 350 | 75 |

Notations are given in the text, Sect. 2.1.

Table 2 Characteristics of the chains

| # | Start | End | Duration, days | Lead time, months | Distance from a strong earthquake, km | Number of earthquakes, $k$ | Maximal distance, $l$, km | Largest magnitude | Area of the R-vicinity, $10^3$ km$^2$ |
|---|---|---|---|---|---|---|---|---|---|
| 1 | 1969.07.16 | 1969.10.03 | 80 | | | 17 | 499 | 5.3 | 150 |
| 2 | 1969.10.15 | 1969.11.19 | 35 | | | 12 | 485 | 5.6 | 113 |
| 3 | 1973.08.26 | 1973.10.17 | 53 | | | 13 | 381 | 4.5 | 150 |
| **4** | **1977.06.03** | **1977.08.01** | **60** | | | 11 | 377 | **4.7** | **104** |
| 5 | 1984.09.07 | 1984.10.26 | 49 | | | 9 | 408 | 4.6 | 90 |
| 6 | 1986.07.08 | 1986.07.20 | 12 | | | 10 | 543 | 5.9 | 122 |
| 7 | 1989.12.24 | 1990.02.04 | 41 | | | 8 | 373 | 5.7 | 101 |
| **8** | **1992.03.27** | **1992.05.08** | **42** | **1.7** | **29** | 17 | 635 | **6.1** | **161** |
| *1992.06.28: Landers earthquake, M=7.6* | | | | | | | | | |
| **9** | **1999.02.19** | **1999.06.01** | **102** | **4.6** | **60** | 11 | 380 | **4.9** | **98** |
| *1999.10.16: Hector Mine earthquake, M=7.4* | | | | | | | | | |

Chains recognized as "precursory" by *RDP* analysis (Sect. 3) are shown in bold. Chain #4 would trigger in prediction a false alarm, chains ##8 and 9 would trigger correct alarms.

As we see in Figs. 2 (two panels in the bottom row) and Fig. 3, only the two last chains (## 8, 9) might be regarded as the local short - time precursors to the Landers and Hector Mine earthquakes: short-term - because they emerge with the short-term lead times (respectively 1.7 and 4.6 months); and local – because the target earthquakes occur in their vicinities. However, the other seven chains, if used as precursors, would give false alarms. To reduce their number we introduce the *RDP* analysis.

### 3. Precursory chains.

**3.1. Hypothesis.** We hypothesize that *a precursory chain (as opposed to a chain giving a false alarm) is preceded by the local intermediate-term precursors formed in the chain's R-vicinity*. This vicinity is not known, until the chain is formed, and its shape might be rather complicated (see Fig. 2). To overcome that impasse we introduce the two-step *RDP* analysis schematically illustrated in Fig. 4.

(*i*) *Search for the chains and determination of their R vicinities* (Sect. 2). Each chain is regarded as a "candidate" for a short-term precursor.

(*ii*) *Search for the local intermediate-term patterns* in the *R*-vicinities of each chain. They are looked for within *T* years before the chain; *T* is an adjustable numerical parameter. If (and only if) such patterns are detected, we regard this chain as a short-term precursor; in prediction it would start a short-term alarm.

To complete that description we have to specify intermediate-term patterns used at the second step.

**3.2. Definitions.** We use the *pattern Σ* which reflects premonitory rise of seismic activity. This pattern, introduced in (Keilis-Borok et al, 1964), is successfully used in different prediction algorithms, alone or in combination with other patterns (Keilis-Borok, 1990, 1996, 2000; Keilis-Borok et al, 1999, 2002; Kossobokov et al, 1995, 2003; Rotwain et al, 1999). It is defined as a premonitory increase of the total area of the earthquake sources. Emergence of this pattern is captured by the function Σ(*t*) defined in a sliding time-window (Keilis-Borok et al, 1964):

$$\Sigma(t/s, B) = \sum_i 10^{Bm_i}, \ m_i \geq m_{min}; \ t-s < t_i \leq t$$

Summation is taken over all main shocks within the time window (*t-s*, t) in the *R*-vicinity of the chain. We take $B \sim 1$, so that the sum is coarsely proportional to the total area of the fault breaks in the earthquakes' sources (Keilis-Borok, 2002); with $B=0$ this sum is the number of earthquakes, with $B = 3/2$ it is proportional to their total energy. The emergence of pattern *Σ* is identified by condition $\Sigma(t) \geq \Sigma_0$; this threshold depends on the magnitude of target earthquakes. In previous applications cited above pattern *Σ* was used as non-local one. We renormalize its numerical parameters to make it local.

**3.3. Data analysis.** We detected precursory chains and determined their *R*-vicinities (Sect. 2). In each vicinity we computed the function Σ(*t*) within time interval *T* = 5 years and summation interval *s* = 6 months. We identified as precursory three chains preceded by largest peaks of Σ(*t*); they can be recognized with the threshold $\Sigma_0 = 10^{6.7}$. Table 2 shows these chains in bold. As we see, identification of the first chain, in 1977, is wrong; in prediction it would give a false alarm. Identification of two other chains, in 1992 and 1999, is correct; each is followed by a target earthquake within few months. The same chains would the selected with the tenfold smaller time interval, *T* = 6 months. The corresponding threshold is $\Sigma_0 = 10^{5.4}$; its value is smaller since smaller number of earthquakes is included in summation.

**3.4. Hypothetical prediction algorithm.** It remains to define alarms triggered by that precursor. This is a final step in transition from a precursor to algorithmic prediction. We adapt the standard general scheme of prediction algorithms, widely used in intermediate-term earthquakes prediction and many other problems (Keilis-Borok, 2002; Kossobokov et al, 1995, and references therein):

(*i*) Prediction is targeted at the main shocks with magnitude M or more; usually the magnitude intervals (*M*, *M*+1) are considered separately.

(*ii*) When a precursory chain is detected, a short-term alarm is triggered. It predicts a target earthquake in *R*-vicinity of the chain, within time interval ($t_e$, $t_e + \tau$); here $t_e$ is the moment when chain emerged, $\tau$ is a numerical parameter (duration of alarm). Results of the data analysis suggest to take $\tau = 6$ months.

*Possible outcomes* of such prediction are illustrated in Fig. 5. Probabilistic component of prediction is represented by the total time-space covered by alarms and probabilities of false alarms and failures to predict (Molchan, 2003).

### 4. Discussion.

**4.1. Summary.** This paper introduces *RDP* analysis in the study of selforganization of seismicity, culminated by a strong earthquake. Precursors with different lead times are considered in reverse order of their appearance. First we detect the candidates for short –term precursors; in our case those are the chains of small earthquakes capturing the rise of earthquake correlation range. A chain determines its narrow vicinity where we look for the local intermediate – term precursor(s), pattern $\Sigma$ in our case. Its presence in turn indicates the precursory chains. We describe *RDP* with detecting precursory chains months before two prominent California earthquakes, Landers, 1992 and Hector Mine, 1999, well isolated in time and space from other comparable earthquakes in that region.

**4.2. Methodological advantage of RDP.** The opposite (direct) analysis would start with detection of the intermediate-term patterns hidden in the background seismicity. Almost all of them, known so far, are not local, pattern $\Sigma$ included. They emerge in the areas whose linear size is up to 10 times larger than the source of the incipient target earthquake (Bowman et al, 1998; Keilis-Borok et al, 2003); some patterns - even up to 100 times larger (Press et al, 1995; Romanowicz, 1993). We have found here renormalized pattern $\Sigma$ that became local: it emerges in the same narrow area (R-vicinity of the chain), where epicenter of a target earthquake lies. As we see in Fig. 2, the shape of that area might be rather complex, and its size - diverse. To find this area by trying different shapes, sizes, and locations is not realistic. Reverse analysis resolves this impasse, indicating a limited number of chains to consider.

**4.3. Physical interpretation.** *RDP* seems to be a promising general approach to prediction of critical transitions in non-linear systems: it identifies a rare small-scale phenomenon that carries a memory of the larger scale history of the system. At the same time this approach has a natural earth-specific explanation: it follows from the concept that strong earthquake is a result of a lasting large-scale process whose different stages involve different parts of the fault network. Earthquakes in the chain mark the part of the fault network that has started to move in unison months before a target earthquake. Pattern $\Sigma$ indicates that this synchronization started much earlier, albeit expressed in a more subtle form. A similar step-by-step escalation of instability was observed in direct analysis: by algorithms M8&MSc (Kossobokov et al, 2003), and by some other algorithms (Aki, 2003; Shebalin et al, 2000; Keilis-Borok et al, 2003).

Both the chains and the peaks of $\Sigma$ are sporadic short-lived phenomena not necessarily reflecting the steady trends of seismicity. This is typical for all premonitory patterns of that family (Keilis-Borok, 2002; Kossobokov et al, 2003). Probably, both patterns are the symptoms but not causes of a strong earthquake: they signal its approach but do not trigger it. Similarly sporadic are many observed precursors to other critical phenomena, e.g. economic recessions (Keilis-Borok et al, 2000).

### 4.4. Implications for earthquake prediction.

-- *RDP* analysis seems worth exploring further on a larger set of target earthquakes. So far it was explored in the following regions: Southern and Central California, 1965-2003, with 9 target earthquakes, M ≥ 6.4.; Japan, 1975-2003, with 12 target earthquakes, M ≥ 7.0; and the Eastern Mediterranean, 1980-2003, with 2 target earthquakes, M = 7.3 and 6.9. For each region numerical parameters have been renormalized to corresponding magnitude thresholds; and all major types of intermediate-term patterns have been considered. The results are highly encouraging (Jin et al, 2003; Shebalin et al, 2003a, b): they are described in more detail in (Shebalin et al, 2003b). In Southern and Central California only one of nine target earthquakes is missed by prediction, with two false alarms. In Japan one of 12 target earthquakes is missed, with seven false alarms; in the Eastern Mediterranean neither target earthquake is missed, with no false alarms.

-- We have the first case of advance prediction: precursory chain reported in advance of a subsequent Tokachi-oki earthquake near Hokkaido island, Japan (Sept. 25, 2003, M = 8.1) (Shebalin et. al, 2003a, b).

-- "Pre-chain" precursors might emerge with a short lead time too.

-- It seems natural to apply the *RDP* analysis to earthquake precursors, expressed in other fields. First positive results are obtained with precursors gauging interaction between the ductile and brittle layers of the crust (Aki, 2003; Jin et al, 2003; Shebalin et al, 2003a). Other promising applications include electromagnetic fields (Uyeda et al, 2002), fluid regime (Keilis-Borok, 1990; Ma et al, 1990), GPS, InSAR, etc.

-- There are no reasons not to explore *RDP* analysis for prediction of critical phenomena in other hierarchical systems, geotechnical, and even soci-economic ones. Similar qualitative approach is routinely used in medicine, criminology, etc.

-- We detect intermediate-term patterns only after a chain has emerged so that its vicinity can be determined; this is too late to declare an intermediate-term alarm. Accordingly, our results concern only short-term prediction.

-- However accurate the short-term prediction would be it will *not* render unnecessary the predictions with a longer lead time. One can find in seismological literature a reappearing mistake: that only short-term prediction is practically useful. Actually, protection from earthquakes requires a hierarchy of preparedness measures, from building codes, insurance, and issuing bonds, to reinforcement of high risk objects, to red alert. It takes different time, from decades, to years, to seconds to undertake different measures. Accordingly, while short-term prediction is currently the most challenging part of prediction research, earthquake preparedness requires *all* stages of prediction (Keilis-Borok, 2002; Molchan, 2003; Kantorovich et al, 1991). Such is the case in preparedness to all disasters, war included.

### 4.5. Questions arising.

-- We considered only one short-term precursor – a chain of earthquakes - and one intermediate-term one - the pattern Σ. In subsequent applications (Shebalin et al, 2003b), all major types of intermediate-term seismicity patterns have been used with similar renormalization. The question arises which set of precursors provides the optimal prediction strategy, as defined for example in (Molchan, 2003; Zaliapin et al, 2003b)

-- It is not yet clear how to make the scaling of *RDP* analysis self-adapting to the regional seismic regime, e. g. to parameters of the Gutenberg-Richter relation.

-- Earthquake precursors emerge with the broader range of the lead times than considered here. They are divided, albeit fuzzily, into *long- term (tens of years)* $\Rightarrow$ *intermediate-term (years)* $\Rightarrow$ *short-term (months) and* $\Rightarrow$ *immediate (days or less)*. The question arises how to apply *RDP* analysis to the whole sequence or to its different parts.

***Summing up***, the *RDP* approach seems to open new possibilities in the quest for the short-term prediction – "the Holy Grail of the earthquake science" according to the US National Reseach Council (Jordan, 2003). We hope that this study sets up a base for further development of this approach in the intertwined problems of earthquake prediction, fundamental understanding dynamics of the lithosphere, and non-linear dynamics.

**Acknowledgements**

This study was supported by the 21st Century Collaborative Activity Award for Studying Complex Systems, granted by the James S. McDonnell Foundation (project "Understanding and Prediction of Critical Transitions in Complex Systems").

Exceedingly useful have been insightful criticism by K. Aki, G. Barenblatt, M. Ghil, R. Mehlman, A. Soloviev, J. Vidale, I. Zaliapin and discussions at the Seventh Workshop on Non-Linear Dynamics and Earthquake Prediction (Trieste, 2003). We are grateful to D. Shatto and V. Ewing for patient and competent help in preparation of the manuscript.

# Figure Captions

Fig. 1. Territory considered.
Stars mark large earthquakes, targeted for prediction. Dots show background seismicity for the time considered (1965-2003): epicenters of earthquakes with magnitude $m \geq 3$ with aftershocks eliminated. Dashed line is used for time-distance projection of epicenters shown in Fig. 3 below.

Fig. 2. Maps of the chains.
Detected chains are shown in separate boxes. Circles show epicenters of earthquakes in a chain; their size is proportional to magnitude. The shadowed areas show $R$-vicinities of the chains. Dates of the beginning and of the end of a chain are given at the top of each box. Three chains (1977, 1992, and 1999) shown in bold are identified as precursory ones. The first chain gives a false alarm; two other chains are followed within few months by target earthquakes, Landers and Hector Mine. Other notations are the same as in Fig.1.

Fig. 3. Chains and strong earthquakes on the time-distance plain.
Distance is counted along the dashed line shown in Fig. 1. Filled and open circles show the chains identified respectively as precursory and non-precursory ones. Other notations are the same as in Fig. 1.

Fig. 4. Schematic illustration of the Reverse Detection of Precursors (*RDP*).
Star is a target earthquake. Light blue circles show earthquakes forming the chain. The rectangle shows symbolically the "R-vicinity" of the chain (the "N. Wiener sausage"). It is the time-space where pattern $\Sigma$ is looked for; its presence indicates a precursory chain. Rectangle on the top shows the map of the chain (yellow) and the source of target earthquake (red). The chains are detected first, although they emerge after the pattern $\Sigma$. Note how a narrow chain determines a much larger time interval where a pattern $\Sigma$ is looked for. Pink area shows the time-space covered by an alarm: within $\tau$ months after precursory chain a target earthquake is expected in its $R$-vicinity.

Fig. 5. Possible outcomes of prediction. Stars mark epicenters of strong earthquakes, targeted by prediction. A box represents the time - space covered by an alarm. A prediction is correct if a strong earthquake occurs within an alarm. Otherwise, this is a false alarm. Failure to predict is the case when a strong earthquake occurs outside of an alarm. Probabilistic component of prediction is represented by the total time-space covered by alarms and probabilities of false alarms and failures to predict.

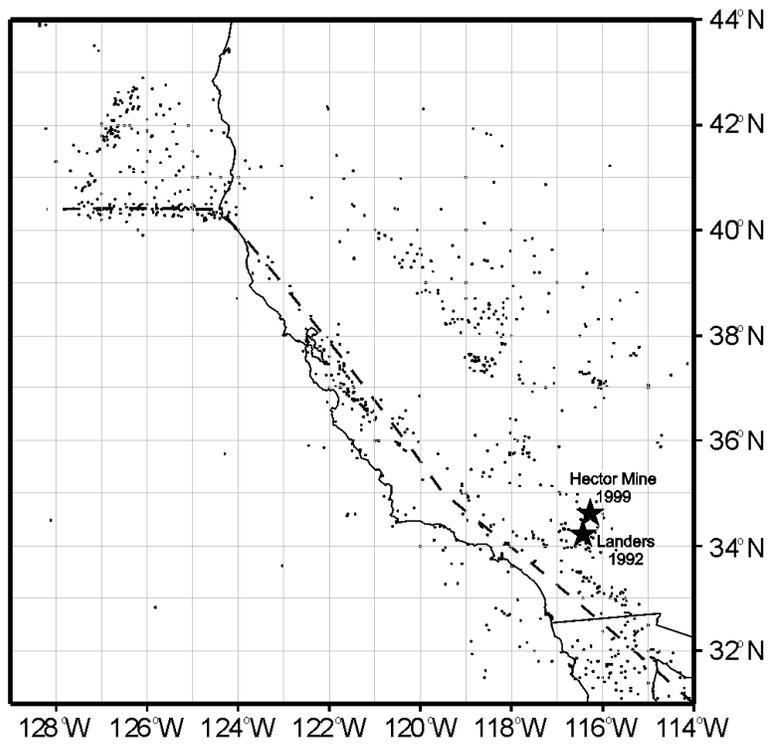

**Figure 1**

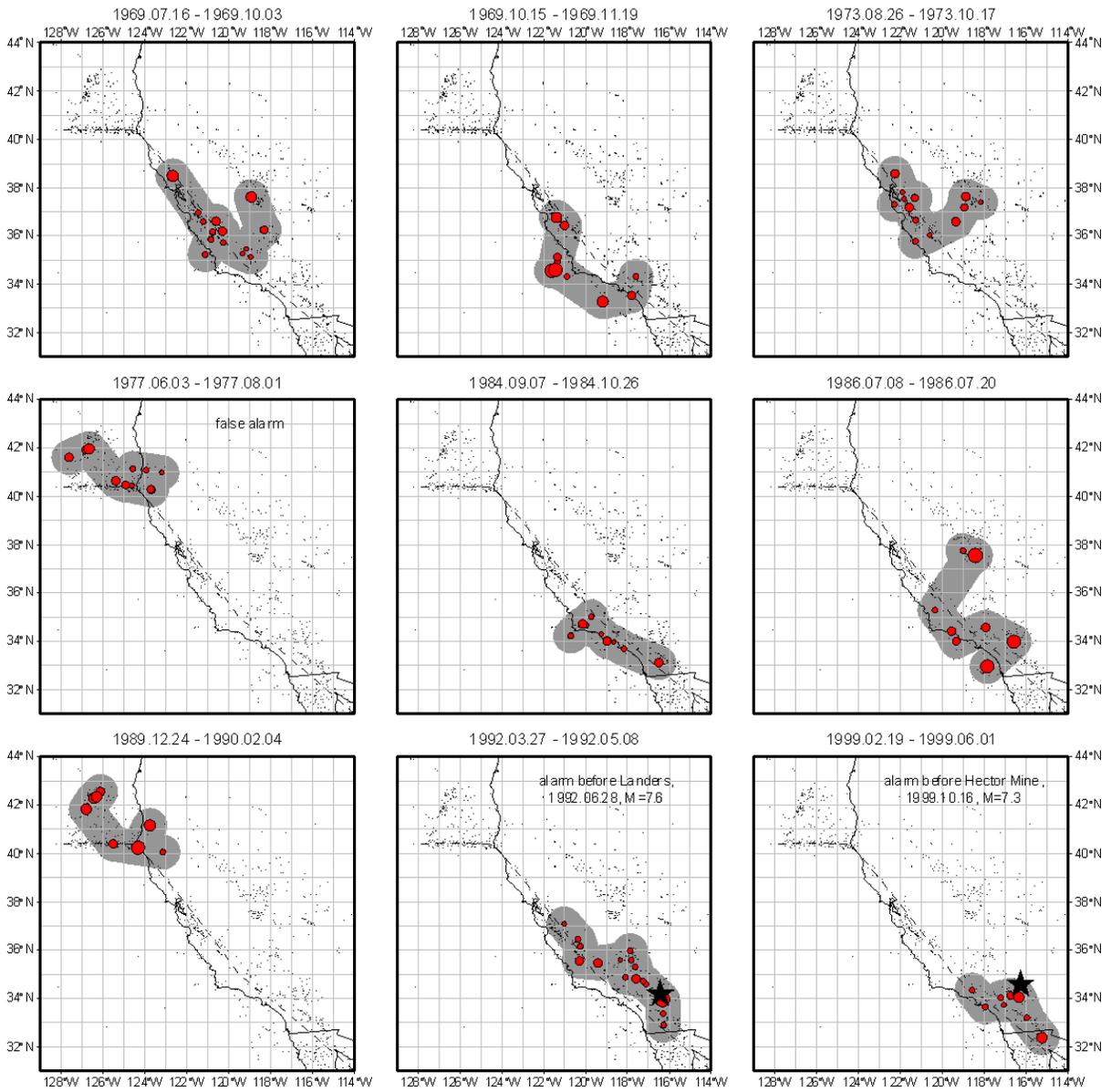

**Figure 2**

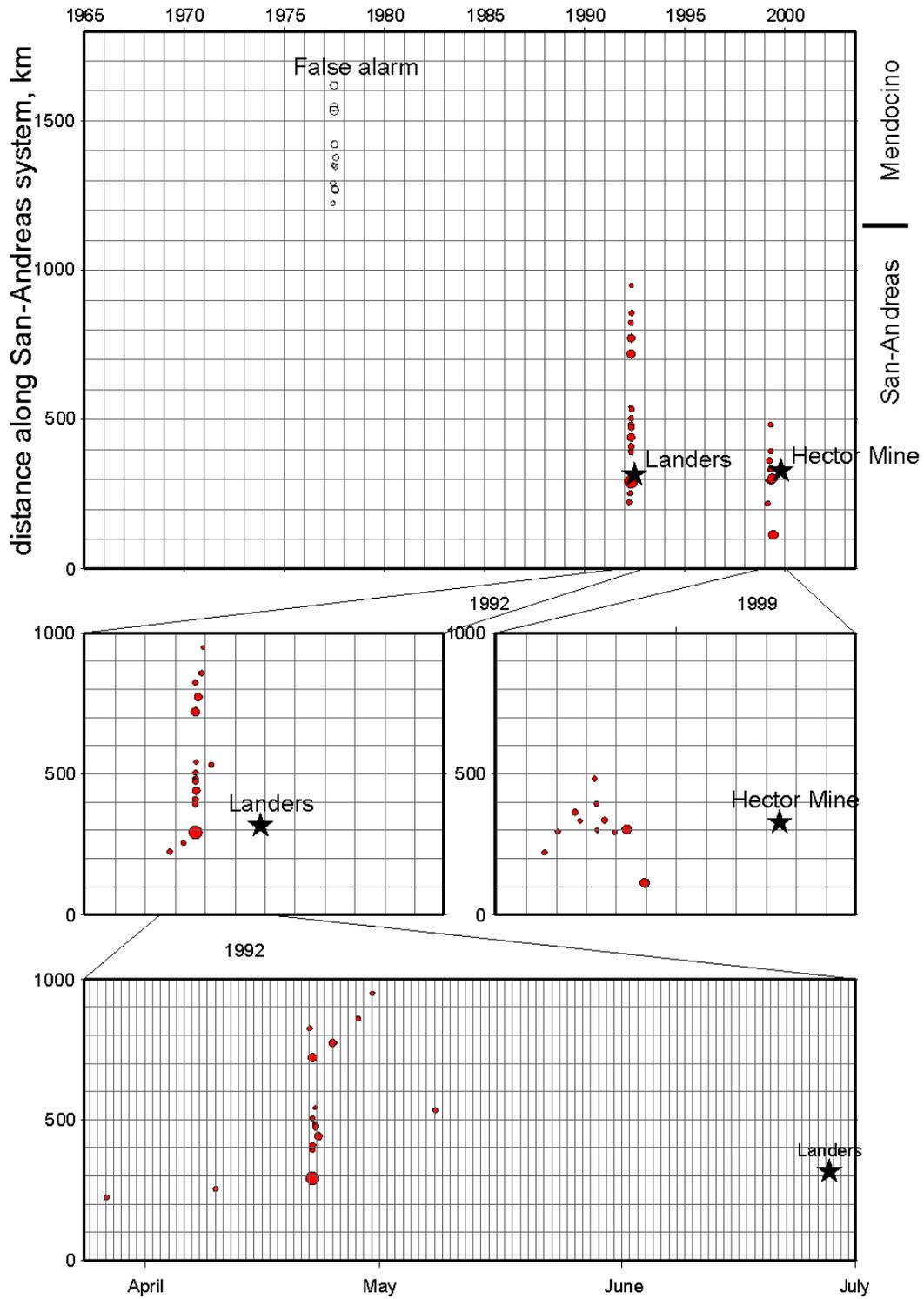

**Figure 3**

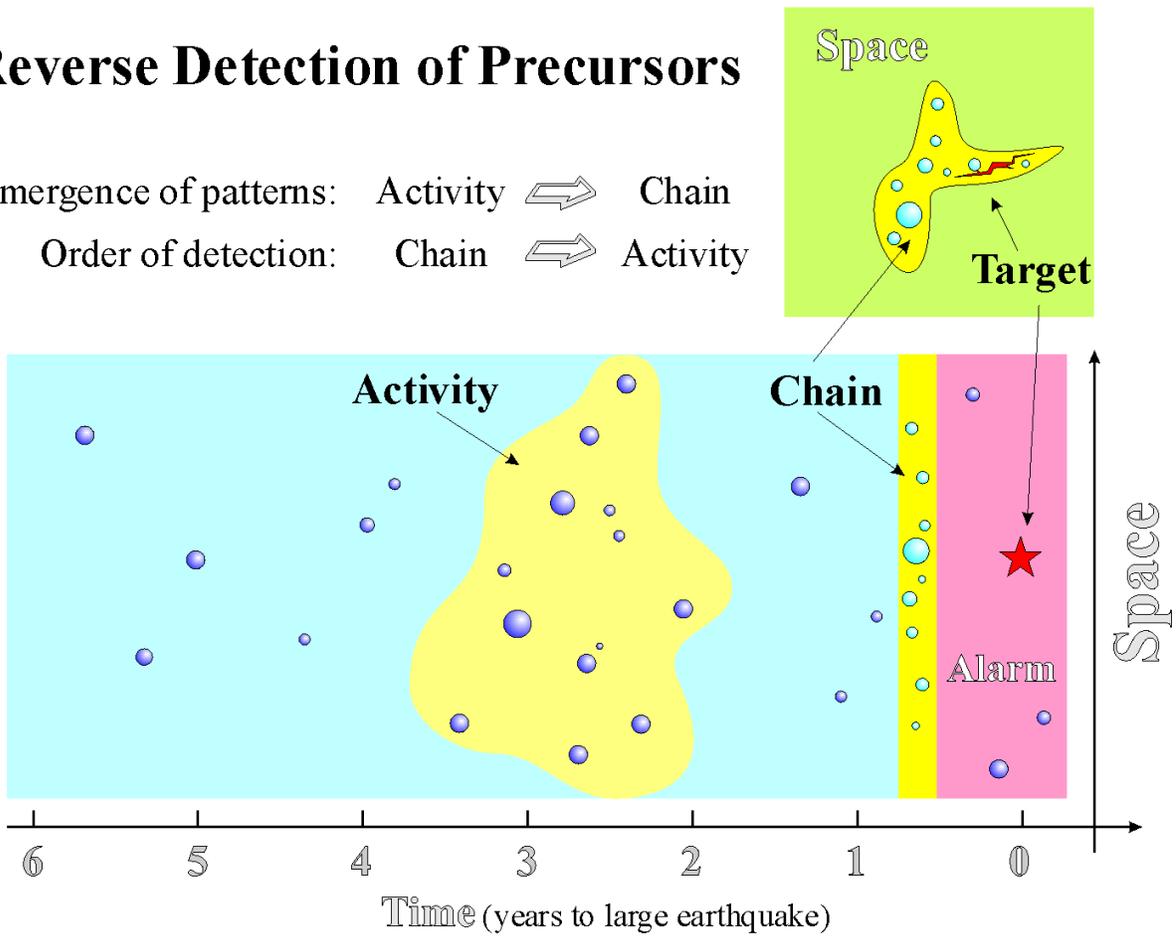

**Figure 4**

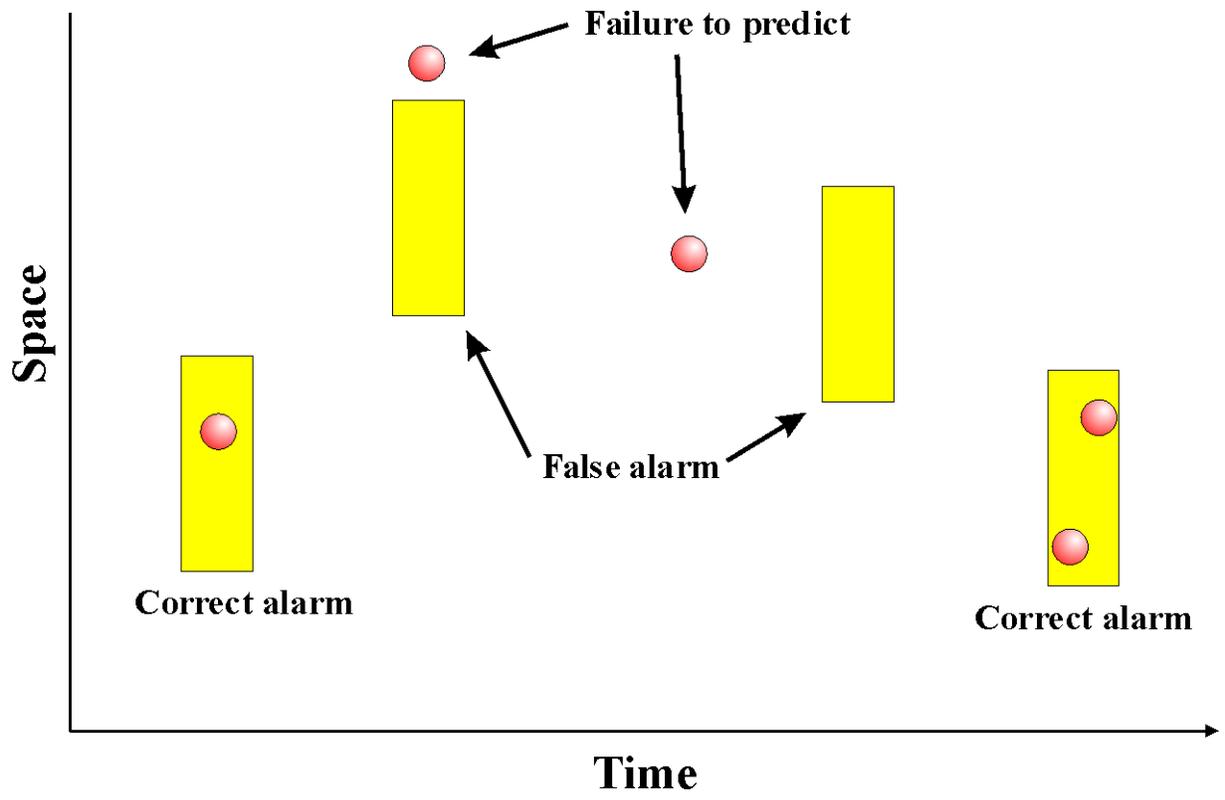

**Figure 5**